\documentclass[english]{article}
\usepackage[T1]{fontenc}
\usepackage[utf8]{inputenc}
\usepackage{geometry}
\usepackage{xcolor}
\usepackage{babel}
\usepackage{float}
\usepackage{amsmath}
\usepackage{amssymb}
\usepackage{graphicx}
\PassOptionsToPackage{normalem}{ulem}
\usepackage{ulem}
\usepackage[unicode=true,pdfusetitle,
 bookmarks=true,bookmarksnumbered=false,bookmarksopen=false,
 breaklinks=false,pdfborder={0 0 0},pdfborderstyle={},backref=false,colorlinks=true]
 {hyperref}
\hypersetup{
 linkcolor=blue,citecolor=blue,urlcolor=blue}

\makeatletter

\providecolor{lyxadded}{rgb}{0,0,1}
\providecolor{lyxdeleted}{rgb}{1,0,0}

\DeclareRobustCommand{\lyxsout}[1]{\ifx\\#1\else\sout{#1}\fi}

\newcommand{\lyxaddress}[1]{
	\par {\raggedright #1
	\vspace{1.4em}
	\noindent\par}
}

\makeatother

\begin{document}
\title{Boundary state bootstrap and asymptotic overlaps in AdS/dCFT}
\author{Tamas Gombor, Zoltan Bajnok}
\maketitle

\lyxaddress{\begin{center}
\emph{Wigner Research Centre for Physics}\\
\emph{Konkoly-Thege Miklós u. 29-33, 1121 Budapest , Hungary}\\
\par\end{center}}
\begin{abstract}
We formulate and close the boundary state bootstrap for factorizing
K-matrices in AdS/CFT. We found that there are no boundary degrees
of freedom in the boundary bound states, merely the boundary parameters
are shifted. We use this family of boundary bound states to describe
the D3-D5 system for higher dimensional matrix product states and
provide their asymptotic overlap formulas. In doing so we generalize
the nesting for overlaps of matrix product states and Bethe states.
\end{abstract}

\section{Introduction}

Boundaries always played significant roles in integrable two dimensional
systems \cite{Ghoshal:1993tm}. When they are placed in space they
provide boundary conditions for the fields, leading to reflection
factors for their particle like excitations. When they are placed
in time they serve as initial or final states which create or annihilate
these particles. Integrable boundary states attracted considerable
interest recently both in non-equilibrium statistical physics and
the AdS/dCFT correspondence. They showed up in integrable quench problems
\cite{Caux:2013ra,Piroli:2018ksf,Piroli:2018don,Pozsgay:2018dzs}
and in the calculation of one point functions in the maximally supersymmetric
gauge theory with a codimension one defect \cite{deLeeuw:2007akd,deLeeuw:2015hxa,Buhl-Mortensen:2015gfd,deLeeuw:2016ofj,Buhl-Mortensen:2016pxs,deLeeuw:2016umh,Buhl-Mortensen:2017ind,deLeeuw:2018mkd,deLeeuw:2019ebw,Linardopoulos:2020jck,Kristjansen:2020mhn}.
In both cases the key object is the overlap of the finite volume boundary
state with finite volume multi-particle states. Although most of the
applications focused on the overlap of the ground state boundary,
recently it was observed in \cite{Komatsu:2020sup} that the description
of the higher dimensional matrix product states (MPS) in the D3-D5
system requires to incorporate the boundary bound states, too. In
\cite{Komatsu:2020sup} the authors focused on the diagonal $\mathfrak{su}(2)$
sub-sector of the theory. Our aim is to extend the boundary bootstrap
procedure for the whole theory including the full matrix structure.
In doing so we develop the bootstrap procedure for boundary states.

Integrable boundary states can be equivalently described by the K-matrix,
which encodes how pairs of particles are annihilated. This K-matrix
is in one-to-one correspondence with the one-particle reflection matrix
in the mirror theory. The boundary Yang-Baxter equation (BYBE) of
the mirror theory is equivalent to the KYBE of the original theory,
which is the consistency equation of the K-matrix \cite{Gombor:2020kgu}.
Much is known about the boundary bootstrap, when poles of the reflection
factor are explained either by boundary bound states or by some boundary
Coleman-Thun diagrams \cite{Dorey:1998kt,Mattsson:2000aw,Bajnok:2001px,Bajnok:2001py,Bajnok:2003dj}.
If the theory and the mirror theory is not equivalent, as in the AdS/CFT
correspondence, then the boundary bootstrap for the reflection factor
is not equivalent to the boundary bootstrap for the boundary state.
This \emph{boundary state bootstrap}, which is equivalent to the boundary
bootstrap for the mirror reflection factor, can be formulated more
intuitively in the original theory. It amounts to explain the poles
of the K-matrix in terms of excited boundary states and Coleman-Thun
diagrams, diagrams with on-shell propagating particles with energy
and momentum preserving point-like interactions \cite{Bajnok:2003dj}.
Our aim is to carry out the boundary state bootstrap for the K-matrix
appearing in the D3-D5 system of the AdS/dCFT correspondence and extend
the result of \cite{Komatsu:2020sup,Gombor:2020kgu}. In doing so
we need the full excitation K-matrix of the D3-D5 system at tree level.
By generalizing the nesting methods of \cite{Gombor:2020kgu} for
matrix product states we calculate these K-matrices, which, together
with the all loop bound state K-matrices, lead to the complete asymptotic
overlaps.

The paper is organized as follows: In section 2 we formulate and close
the boundary bootstrap for the factorizing K-matrices we found in
\cite{Gombor:2020kgu}. We then generalize the nesting for matrix
product states' overlaps and extract the tree level excitation K-matrices
by explicit calculations in section 3. In the end of section 3 we propose our
formulas for the full asymptotic overlap including all sectors. Finally,
we conclude in section 4.

\section{Boundary state bootstrap}

In the $AdS_{5}/CFT_{4}$ correspondence the scattering matrix factorizes
into the product of two $\mathfrak{su}(2\vert2)$ invariant scattering
matrices \cite{Beisert:2010jr}:
\begin{equation}
\mathbb{S}(p_{1},p_{2})=S_{0}(p_{1},p_{2})S(p_{1},p_{2})\otimes S(p_{1},p_{2})
\end{equation}
 In our previous paper we have determined the most general factorizing
solutions
\begin{equation}
\mathbb{K}(p)=K_{0}(p)K(p)\otimes K(p)
\end{equation}
of the KYBE without inner degrees of freedom \cite{Gombor:2020kgu}.
We found that they all have centrally extended $\mathfrak{osp}(2\vert2)_{c}$
residual symmetry and the various solutions were characterized by
how this symmetry can be embedded into the centrally extended $\mathfrak{su}(2\vert2)_{c}$.
There were two classes depending on whether the unbroken $\mathfrak{su}(2)$
symmetry was in the Lorentz or in the $R$-symmetry part. As these
two cases are quite analogous we focused on the solution, which preserves
the Lorentz symmetry and can be written as

\begin{equation}
K(p)=\left(\begin{array}{cccc}
k_{1} & k_{2}+e(p) & 0 & 0\\
k_{2}-e(p) & k_{4} & 0 & 0\\
0 & 0 & 0 & f(p)\\
0 & 0 & -f(p) & 0
\end{array}\right)\quad;\qquad\begin{array}{c}
e(p)=-i\frac{x_{s}^{2}x^{-}+x^{+}}{x_{s}(1+x^{+}x^{-})}\\
f(p)=i\sqrt{\frac{x^{-}}{x^{+}}}\frac{(x^{+})^{2}-x_{s}^{2}}{x_{s}(1+x^{+}x^{-})}\\
k_{1}k_{4}-k_{2}^{2}=1
\end{array}\label{eq:Kmatrix}
\end{equation}
where $x^{\pm}\equiv x^{\pm}(p)$ are the standard parameters of the
one-particle representations. This parametrization is slightly different
from \cite{Gombor:2020kgu} by rescaling $k_{i}$, which now parameterize
the orientation of the bosonic part of the $\mathfrak{osp}(2\vert2)_{c}\subset\mathfrak{su}(2\vert2)_{c}$
embedding, while the parameter\footnote{We introduced $x_{s}$ to conform with the notation of \cite{Komatsu:2020sup}.
It is related to $s$ in \cite{Gombor:2020kgu} as $s=-ix_{s}$.} $x_{s}$ is responsible for the fermionic orientation $\tilde{\mathbb{Q}}_{\alpha}^{\:\:a}=\mathbb{Q}_{\alpha}^{\:\:a}+ix_{s}^{-1}\epsilon_{\alpha\beta}\sigma_{1}^{ab}\mathbb{Q}_{b}^{\dagger\:\beta}$,
with $\sigma_{1}$ being the first Pauli matrix. Our normalization
is also different from \cite{Gombor:2020kgu}, as we removed the factor
$\frac{-ix_{s}(1+x^{+}x^{-})}{(x^{+})^{2}-x_{s}^{2}}$ and put it
directly in to $K_{0}(p)$. This choice is more useful for bootstrap
purposes and the boundary crossing equation takes also a very simple
form 
\begin{equation}
K_{0}(p)=S_{0}(p,-p)K_{0}(p)\quad;\qquad S_{0}(p_{1},p_{2})=\frac{x_{1}^{-}}{x_{1}^{+}}\frac{x_{2}^{+}}{x_{2}^{-}}\frac{x_{1}^{+}-x_{2}^{-}}{x_{1}^{-}-x_{2}^{+}}\frac{1-\frac{1}{x_{1}^{+}x_{2}^{-}}}{1-\frac{1}{x_{1}^{-}x_{2}^{+}}}\sigma(p_{1},p_{2})^{2}\label{eq:Kcrossing}
\end{equation}
where $S_{0}(p_{1},p_{2})$ is the scalar factor in the $\mathfrak{su}(2)$
sector. The $K$-matrix satisfies the KYBE: 
\begin{equation}
K_{23}(p_{2})K_{14}(p_{1})S_{13}(p_{1},-p_{2})S_{13}(p_{1},p_{2})=K_{14}(p_{1})K_{23}(p_{2})S_{24}(p_{2},-p_{1})S_{34}(-p_{2},-p_{1})\label{eq:KYBE}
\end{equation}
where the subscripts indicate in which representation spaces the operators
act $(p_{1},p_{2},-p_{2},-p_{1})$. The reflection factor of the mirror
theory defined by $R(p(z))=CK(p(\frac{\omega_{2}}{2}-z))$ satisfies
the boundary Yang-Baxter equation \cite{Gombor:2020kgu}. Here $z$
is the torus rapidity parameter, $C$ is the charge conjugation matrix
and $\omega_{2}$ is the crossing parameter. Unitarity of the mirror
reflection factor implies the following equation for the scalar factor
\begin{equation}
K_{0}(p(z+\omega))K_{0}(p(z))=\frac{x_{s}^{4}(1+x^{+}x^{-})^{4}}{((x^{+})^{2}-x_{s}^{2})^{2}(1-(x^{-})^{2}x_{s}^{2})^{2}}\label{eq:unitarity}
\end{equation}
The minimal solution to the crossing (\ref{eq:Kcrossing}) and unitarity
equation (\ref{eq:unitarity}) was found in the D3-D5 setting in \cite{Komatsu:2020sup}.
Our conventions are related to those by $x^{\pm}\leftrightarrow x^{\mp}$
which originate from the different definition of the scattering matrix\footnote{Part of the literature uses the $S$-matrix \cite{Arutyunov:2009ga},
while some other part the inverse of it~\cite{Arutyunov:2007tc}.
This can be easily pinpointed by how the S-matrix appears in the Bethe
ansatz equations. Accordingly, the physical domain of the rapidity
is also exchanged in the two conventions. We use the conventions of
\cite{Arutyunov:2007tc}.}. We also have a slightly different normalization of the K-matrix.
In this paper we focus on the bootstrap and the matrix part of the
solution together with the corresponding overlaps, so we just point
out that $K_{0}(z)$ can be obtained from that of \cite{Komatsu:2020sup}
by the corresponding changes. In order to keep the discussion in a
more general level we assume that the scalar factor has poles at $x^{+}=\pm x_{s}$,
which come from the removed factor $\frac{-ix_{s}(1+x^{+}x^{-})}{(x^{+})^{2}-x_{s}^{2}}$.
Following \cite{Komatsu:2020sup} one might regard $x_{s}$ as the
$x$ parameter of a boundary rapidity $is,$ which satisfies: $x_{s}+x_{s}^{-1}=\frac{is}{g}.$
The poles signal boundary bound states and in the following we calculate
the K-matrices of the corresponding excited boundary state.

\subsection{Boundary bound state K-matrix\label{subsec:Boundary-boundstate-K-matrix}}

\begin{figure}[H]
\begin{centering}
\includegraphics[width=6cm]{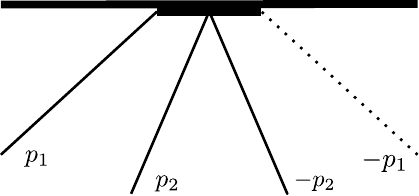}~~~~~~~~\includegraphics[width=6cm]{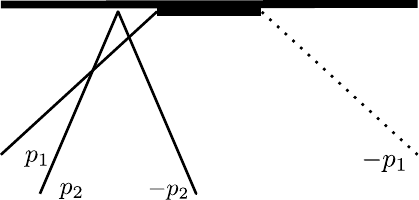}
\par\end{centering}
\caption{Boundary bootstrap for the K-matrix (boundary state). Trajectories
can be shifted allowing to calculate the bound state K-matrix in terms
of the groundstate one and scattering matrices.}

\label{fig:bootstrap}
\end{figure}

The bootstrap principle tells us that the $K$-matrix of the boundary
bound state $\bar{K}(p)$ can be calculated from the ground state
one by shifting the trajectories of the particles as show on Figure
\ref{fig:bootstrap}:
\begin{equation}
\bar{K}_{23}(p_{2})\text{res}_{x_{1}^{+}=\pm x_{s}}K_{14}(p_{1})=\text{res}_{x_{1}^{+}=\pm x_{s}}(K_{23}(p_{2})K_{14}(p_{1})S_{13}(p_{1},-p_{2})S_{12}(p_{1},p_{2}))\label{eq:Kbootstrap}
\end{equation}

It is particularly useful to normalize the $\mathfrak{su}(2\vert2)$
scattering matrix as $S_{11}^{11}(p_{1},p_{2})=1$. By performing
the calculation it turns out that the K-matrix at the pole position
is proportional to a one dimensional projector thus the bound state
K-matrix has the same dimension as the original one. This implies
that there is no boundary degree of freedom in the boundary bound
state, similarly to the boundary ground state with which we started.
Due to the bootstrap construction the bound state K-matrix satisfies
the KYBE (\ref{eq:KYBE}) thus has the form \ref{eq:Kmatrix} and
can be described by other constants $\bar{s},\bar{k}_{i}$. On bootstrapping
on the poles $x_{1}^{+}=\pm x_{s}$ we found that, due to our special
normalization, the parameters of the bound state K-matrix can be obtained
from the original one as 
\begin{equation}
s\to\bar{s}=s\mp1\quad;\qquad k_{i}\to\bar{k}_{i}
\end{equation}
and the scalar factor changes as 
\begin{equation}
\bar{K}_{0}(p)=S_{0}(p_{s},p)S_{0}(p_{s},-p)K_{0}(p)\quad;\qquad x^{+}(p_{s})=\pm x_{s}\label{eq:scalarbootstrap}
\end{equation}

Similar calculations can be done for the other $\mathfrak{osp}(2\vert2)_{c}$
residual symmetry, when the unbroken $\mathfrak{su}(2)$ part is the
R-symmetry with the same conclusion, i.e. there is no boundary degrees
of freedom and the parameters change the same way: $s\to\bar{s}=s\mp1$
depending on which pole we bootstrap on and the $\bar{k}$s are not
changed.

\subsection{Symmetry considerations}

These results can be easily understood from symmetry considerations.
Indeed, the fermionic symmetry of the K-matrix can be written formally
as \cite{Gombor:2020kgu} 
\begin{equation}
K(p)\Delta(\mathbb{Q}+x_{s}^{-1}\mathbb{Q}^{\dagger})=0
\end{equation}
The shift in the parameter $s$ comes from the different co products
of $\mathbb{Q}$ and $\mathbb{Q^{\dagger}}$: 
\begin{align}
\Delta(\mathbb{Q}) & =\mathbb{Q}\otimes e^{i\mathbb{P}/4}+e^{-i\mathbb{P}/4}\otimes\mathbb{Q}, & \Delta(\mathbb{Q}^{\dagger}) & =\mathbb{Q}^{\dagger}\otimes e^{-i\mathbb{P}/4}+e^{i\mathbb{P}/4}\otimes\mathbb{Q}^{\dagger}
\end{align}
via the bootstrap procedure, which is obtained through the pole term
of the KYBE at $x^{+}(p_{1})=\pm x_{s}$. Since the KYBE (\ref{eq:KYBE})
is an equation on the product of one particle $\mathfrak{su}(2|2)_{c}$
representations with momenta $\mathcal{V}(p_{1})\otimes\mathcal{V}(p_{2})\otimes\mathcal{V}(-p_{2})\otimes\mathcal{V}(-p_{1})$
the co-product of the supercharges takes the following form 
\begin{align}
\Delta^{(4)}(\mathbb{Q}) & =e^{-i\frac{p_{1}}{4}}\mathbb{Q}\otimes\mathbb{I}\otimes\mathbb{I}\otimes\mathbb{I}+e^{-i\frac{p_{1}}{4}}\mathbb{I}\otimes\mathbb{I}\otimes\mathbb{I}\otimes\mathbb{Q}+e^{-ip_{1}/2}(\mathbb{I}\otimes\Delta(\mathbb{Q})\otimes\mathbb{I})\label{eq:coprod4}\\
 & =e^{-i\frac{p_{1}}{4}}(\mathbb{Q}_{1}+\mathbb{Q}_{4})+e^{-ip_{1}/2}\Delta_{23}(\mathbb{Q}).\nonumber 
\end{align}
This can be decomposed with respect to the particle types as the action
of the charge on the representations with momentum $p_{1}$ and $-p_{1}$
(first two terms) and $e^{-ip_{1}/2}$ times the action of the charge
on the representations with momentum $p_{2}$ and $-p_{2}$ (second
term). In a similar manner the action of $\mathbb{Q}^{\dagger}$ comes
with a factor $e^{ip_{1}/2}$ when acting in the representations of
the second particle:
\begin{equation}
\Delta^{(4)}(\mathbb{Q}^{\dagger})=e^{i\frac{p_{1}}{4}}(\mathbb{Q}_{1}^{\dagger}+\mathbb{Q}_{4}^{\dagger})+e^{ip_{1}/2}\Delta_{23}(\mathbb{Q}^{\dagger}).
\end{equation}
 We now act with $\Delta^{(4)}(\mathbb{Q}+x_{s}^{-1}\mathbb{Q}^{\dagger})$
on the bootstrap equation (\ref{eq:Kbootstrap}): the r.h.s is zero
since the scattering matrices commute with the charges and the boundary
states $K_{14}$ and $K_{23}$ are annihilated by $\Delta(\mathbb{Q}+x_{s}^{-1}\mathbb{Q}^{\dagger})$,
which implies:
\begin{equation}
\left[\bar{K}_{23}(p_{2})\text{res}_{x_{1}^{+}=\pm x_{s}}K_{14}(p_{1})\right]\Delta^{(4)}(\mathbb{Q}+x_{s}^{-1}\mathbb{Q}^{\dagger})=0.
\end{equation}
Using (\ref{eq:coprod4}) and the fact that
\begin{equation}
K_{14}(p_{1})\left(e^{-i\frac{p_{1}}{4}}(\mathbb{Q}_{1}+\mathbb{Q}_{4})+x_{s}^{-1}e^{i\frac{p_{1}}{4}}(\mathbb{Q}_{1}^{\dagger}+\mathbb{Q}_{4}^{\dagger})\right)=0
\end{equation}
we can obtain the following property of the bound state K-matrix:
\begin{equation}
\bar{K}(p_{2})\Delta(\mathbb{Q}e^{-i\frac{p_{1}}{2}}+x_{s}^{-1}e^{i\frac{p_{1}}{2}}\mathbb{Q}^{\dagger})=0
\end{equation}
This shows that the parameter $x_{s}=\pm x_{1}^{+}$ in the bound
state K-matrix is shifted compared to the original one as: 
\begin{equation}
\bar{K}(z)\Delta(\mathbb{Q}+x_{\bar{s}}^{-1}\mathbb{Q}^{\dagger})=0\quad;\qquad x_{\bar{s}}=e^{-ip_{1}}x_{s}=\pm x_{1}^{-}=x_{s\mp1}
\end{equation}

\subsection{Closing the boundary bootstrap}

The newly calculated bound state $K$-matrix $\bar{K}$ with parameter
$\bar{s}$ has the same structure as the ground state one except that
the parameter is changed $s\to\bar{s}$. This signals an another boundary
bound state which we can excite by binding a particle with $x^{+}(p)=\pm x_{\bar{s}}$
to the already excited boundary. The boundary bootstrap formulates
how this procedure goes. The closure of the bootstrap requires that
we explain all poles of all the excited K-matrices. Explanation means
that for each singularity we draw a boundary Coleman-Thun diagram
with on-shell propagating particles. In relativistic theories this
is the Coleman-Norton interpretation of the Landau equations coming
from all loop perturbative calculations which explains the pole singularities
\cite{Bajnok:2003dj}.

In performing a complete analysis of the possible bound states the
proper scalar factor is essential. So from now on we focus only on
the D3-D5 system, where the scalar factor is known \cite{Komatsu:2020sup}.
This scalar factor was fixed by solving the crossing (\ref{eq:Kcrossing})
and unitarity (\ref{eq:unitarity}) equations for $K_{0}$. In our
K-matrix we have a prefactor $(x^{+}-x_{s})(x^{+}+x_{s})$ in the
denominator indicating two possible bound states. However, in the
solution for $K_{0}$ it is replaced by the following structure:
\begin{equation}
\frac{(x^{+}+\frac{1}{x^{+}})(x^{+}+\frac{1}{x^{+}}+x^{-}+\frac{1}{x^{-}})}{2(x^{+}-x_{s})(1-\frac{1}{x^{+}x_{s}})(x^{-}+x_{s})(1+\frac{1}{x^{-}x_{s}})}\times\text{regular}
\end{equation}
It is thus advantageous to introduce the rapidity variable also for
the particle, similarly to the boundary parameter:

\begin{equation}
x^{\pm}(p)+\frac{1}{x^{\pm}(p)}=\frac{u(p)\pm\frac{i}{2}}{g}\quad;\qquad x_{s}+\frac{1}{x_{s}}=\frac{is}{g}
\end{equation}
such that the scalar factor has the form 
\begin{equation}
K_{0}(u)=\frac{u(u+\frac{i}{2})}{(u-i(s-\frac{1}{2}))(u+i(s-\frac{1}{2}))}K_{0}^{\mathrm{reg}}(u)
\end{equation}
This result is the conjugate expression ($x^{\pm}\leftrightarrow x^{\mp}$)
of the one in \cite{Komatsu:2020sup}, coming from the fact that we
use a different convention for the scattering matrix \cite{Arutyunov:2007tc}.
This conjugation, however will not effect the final physical overlaps.
The regular part contains the ratio of the boundary dressing phase
and the bulk dressing phase and will not be relevant for closing the
bootstrap which we perform now\footnote{The boundary dressing phase does not have any pole in the physical
domain, while the singularities of the bulk dressing phase correspond
to bulk bound states which we do not analyze here.}. We start with the boundary ground state. The physical domain of
the $x$ parameters is $\vert x^{\pm}\vert>1$ thus, when we have
a pole in $u$ at $i(s-\frac{1}{2})$, we bootstrap on the pole $x^{+}=x_{s}$
and not on the pole at $x^{+}=\frac{1}{x_{s}}$. This is also true
for all bound states.

\subsubsection{Boundary ground state}

\begin{figure}[H]
\begin{centering}
\includegraphics[width=5cm]{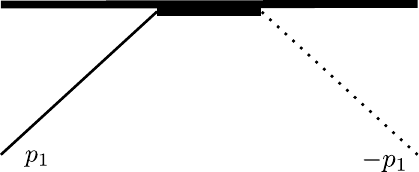}~~~~~~~~\includegraphics[width=4cm]{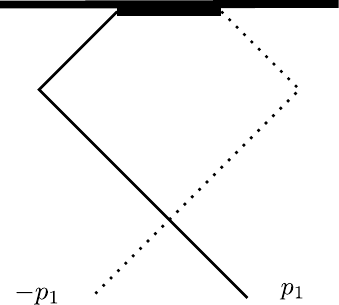}
\par\end{centering}
\caption{Coleman-Thun diagrams for the poles of the ground state K-matrix.
They correspond to a boundary bound state pole in the direct and one
in the crossed diagram indicated by a thicker boundary line.}

\label{fig:boundstate}
\end{figure}

In the K-matrix we have two poles in $K_{0}(u_{1})$. One at $u_{1}=i(s-\frac{1}{2})$
and another one at $u_{1}=-i(s-\frac{1}{2})$. At $u_{1}=i(s-\frac{1}{2})$
the K-matrix is a rank one projector and the pole is explained by
a boundary bound state without any inner degree of freedom as shown
on the left of figure (\ref{fig:boundstate}).

Due to the boundary crossing unitarity for any pole in the K-matrix
at $u_{p}$ there is also a pole at $-u_{p}$. This is similar how
in the bulk bootstrap the crossing invariant scattering matrix has
poles both in the $s$- and also in the $t$- channel. In order to
decide which is the boundary bound state we can analyze the $K$-matrix
at the pole position. Clearly in our case at the pole $u_{1}=i(s-\frac{1}{2})$
we have a one dimensional projector, so it is natural to associate
a boundary bound state to it. At the position $u_{1}=-i(s-\frac{1}{2})$
we have a three dimensional projector as we have poles also in the
$34$ and $43$ matrix elements, but they can be explained by the
diagram on the right of figure (\ref{fig:boundstate}). As $34$ can
scatter into $12$ which creates the bound state at $u_{1}$ we effectively
created the bound state by the 34 process at $-u_{1}$. If, instead,
we had assumed that the three dimensional projector corresponds to
the physical bound state then we would not have been able to use the
crossed channel diagram to explain the one dimensional projector structure
of the other pole. Thus keeping the first option and using the two
diagrams we explained all the ground state poles and proceed with
bootstrapping the K-matrix for the boundary bound state.

\subsubsection{First boundary excited state}

We now use the boundary state bootstrap we described in the beginning
of the section to calculate the bound state K-matrix and the scalar
factor (\ref{eq:Kbootstrap}). This means that a particle with label
$1$ and momentum $x_{1}^{+}=x_{s}$, binds to the boundary. By bootstrapping
on the $u_{1}=i(s-\frac{1}{2})$ pole we obtain the excited boundary
K-matrix, $\bar{K}(u_{2})$ with parameter $\bar{s}=s-1$. The scalar
factor can be bootstrapped as 
\begin{align}
K_{1}(u_{2}) & =K_{0}(u_{2})S_{0}(u_{1},u_{2})S_{0}(u_{1},-u_{2})\\
 & =\frac{u_{2}(u_{2}+\frac{i}{2})}{(u_{2}-i(s-\frac{1}{2}))(u_{2}+i(s-\frac{1}{2}))}\frac{u_{1}-u_{2}+i}{u_{1}-u_{2}-i}\frac{u_{1}+u_{2}+i}{u_{1}+u_{2}-i}K_{1}^{\mathrm{reg}}(u_{2})\nonumber 
\end{align}
Using that $u_{1}=i(s-\frac{1}{2})$ we obtain the excited boundary
state scalar factor 
\begin{equation}
K_{1}(u)=\frac{u(u+\frac{i}{2})}{(u-i(s-\frac{1}{2}))(u+i(s-\frac{1}{2}))}\frac{u-i(s+\frac{1}{2})}{u-i(s+\frac{3}{2})}\frac{u+i(s+\frac{1}{2})}{u+i(s-\frac{3}{2})}K_{1}^{\mathrm{reg}}(u)
\end{equation}
Observe that additionally to the pair of ground state poles we obtained
a new pair shifted by $1$ according to the new $s$ parameter.

\begin{figure}
\begin{centering}
\includegraphics[width=5cm]{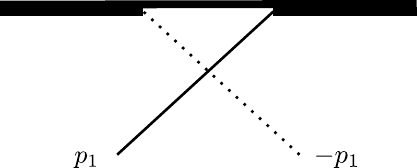}
\par\end{centering}
\caption{Coleman-Thun diagram for excited boundary states signaling a pole
at the ground state excitation position.}

\label{fig:emitabsorb}
\end{figure}

The ground state pole always appears in the excited state and can
be explained by the diagram on figure (\ref{fig:emitabsorb}). There
is a crossed version of this diagram, which explains the pole at $-u_{1}$.
At these positions the scattering matrix is not a projector. This
is not a problem however, since before reaching the boundary the particles
have to scatter on each other allowing poles in all amplitudes. In
particular, the particles $34$ can scatter into $12$ or $21$, which
then can be emitted by the excited boundary and absorbed by the ground
state boundary.

\subsubsection{Second boundary excited state}

The pole at $u=i(s-\frac{3}{2})$ signals a new bound state. Performing
the bootstrap we assume that a particle with rapidity $u_{2}=i(s-\frac{3}{2})$
binds to the already excited boundary. The K-matrix of the newly excited
boundary $\bar{K}(u_{3})$ has a new s-parameter whose rapidity turns
out to be $s-2$ and the scalar factor of a particle with rapidity
$u_{3}$ is
\begin{align}
K_{2}(u_{3}) & =K_{1}(u_{3})S_{0}(u_{2},u_{3})S_{0}(u_{2},-u_{3})\\
 & \sim K_{1}(u_{3})\frac{u_{2}-u_{3}+i}{u_{2}-u_{3}-i}\frac{u_{2}+u_{3}+i}{u_{2}+u_{3}-i}\nonumber 
\end{align}
The numerator of the scattering matrix factors cancel the ground state
bound state poles and we obtain 
\begin{equation}
K_{2}(u)=\frac{u(u+\frac{i}{2})}{(u-i(s-\frac{3}{2}))(u+i(s-\frac{3}{2}))}\frac{u-i(s+\frac{1}{2})}{u-i(s-\frac{5}{2})}\frac{u+i(s+\frac{1}{2})}{u+i(s-\frac{5}{2})}K_{2}^{\mathrm{reg}}(u)
\end{equation}
Clearly we have the pole corresponding to decaying and exciting and
its crossed versions. Fortunately, the pole at $u=i(s-\frac{1}{2})$
disappeared, which would have been very difficult to explain. The
new pole signals a new boundary bound state. We can repeat this procedure
to obtain the generic $n^{th}$ excited boundary state.

\subsubsection{Generic boundary excited state}

The generic $n^{th}$ boundary excited state has a boundary rapidity
parameter $s-n$ and prefactor
\begin{equation}
K_{n}(u)\propto\frac{u(u+\frac{i}{2})}{(u-i(s-n+\frac{1}{2}))(u+i(s-n+\frac{1}{2}))}\frac{u-i(s+\frac{1}{2})}{u-i(s-n-\frac{1}{2})}\frac{u+i(s+\frac{1}{2})}{u+i(s-n-\frac{1}{2})}
\end{equation}
Clearly the first pole is a decay-excitation pole, while the second
one is signaling a new boundary bound state.

\subsubsection{Closing the bootstrap}

For generic parameter $s$ the bootstrap never stops and we have an
infinite family of boundary bound states.

The bootstrap can close if the factor $(u+i(s+\frac{1}{2}))$ in the
numerator cancels the would be bound state pole at $u-i(s-N-\frac{1}{2})$.
This happens when 
\begin{equation}
s=\frac{N}{2}
\end{equation}
where $N$ is an integer. But we have to be careful in this case.
Indeed, if for example $N$ is odd then after several steps we obtain
the denominator $(u-i)u^{2}(u+i)$ where the new boundary bound state
should be at $u=0$. Double pole, however does not correspond to any
bound state. Fortunately, $u$ in the prefactor just transforms the
double pole into a single one which is appropriate for a new boundary
bound state. For even $N$ similar situation appears around the origin
at $(u-\frac{i}{2})^{2}(u+\frac{i}{2})^{2}$ where the should be single
pole is at $u=-\frac{i}{2}$. Clearly the prefactor $u+\frac{i}{2}$
helps again and the bootstrap does not stop around the origin in either
case.

Thus we can conclude that for $s=\frac{N}{2}$ with $N$ integer the
bootstrap is closed and we have $N+1$ boundary bound states with
rapidities $\frac{i}{2}(N,N-2,\dots,2-N,-N)$. For the comparison
with the weak coupling expansion we take $N=2j$ to transform the
spectrum of rapidities into $i(j,j-1,\dots,1-j,j)$. By labeling the
bound state with its parameter $m=(j,j-1,\dots,1-j,j)$ the corresponding
scalar factor is 
\begin{equation}
K_{j,m}(u)=\frac{u(u+\frac{i}{2})}{(u-\frac{(2m-1)i}{2})(u+\frac{(2m-1)i}{2})}\frac{(u-\frac{i(2j+1)}{2})(u+\frac{i(2j+1)}{2})}{(u-\frac{(2m+1)i}{2})(u+\frac{(2m+1)i}{2})}K_{j,m}^{\mathrm{reg}}(u)\label{eq:prefacKm}
\end{equation}
where, in the notations we made explicit the j-dependence, too. In
contrast, the matrix part depends only on $m$ and not on $j$. Let
us denote this matrix, (\ref{eq:Kmatrix}) with $x_{s}\to x_{m}$,
by $K^{(m)}(u)$ such that the full K-matrix becomes
\begin{equation}
\mathbb{K}^{(j,m)}(u)=K_{j,m}(u)K^{(m)}(u)\otimes K^{(m)}(u)\label{eq:Kjm}
\end{equation}

Let us emphasize that this result goes much beyond \cite{Komatsu:2020sup}.
In \cite{Komatsu:2020sup} the authors only calculated the prefactor
(\ref{eq:prefacKm}) and did not consider the matrix part, $K^{(m)}$,
nor explained the other, non-boundstate, poles of the reflection factor
or considered the closure of the bootstrap.

In order to match our results to the AdS/dCFT calculations we perform
a weak coupling expansion.

\subsection{Weak coupling expansion\label{subsec:Weak-coupling-expansion}}

Assuming that $k_{i}$ are coupling independent we can make a weak
coupling expansion of the bound state K-matrices easily. The matrix
part at leading order takes the form

\begin{equation}
\lim_{g\to0}K^{(m)}(u)=\left(\begin{array}{cccc}
k_{1} & k_{2}+\frac{m}{u+\frac{i}{2}} & 0 & 0\\
k_{2}-\frac{m}{u+\frac{i}{2}} & k_{4} & 0 & 0\\
0 & 0 & 0 & 0\\
0 & 0 & 0 & 0
\end{array}\right)
\end{equation}
The weak coupling expansion of the prefactor $K_{j,m}(u)$ is (\ref{eq:prefacKm})
without the regular part. This will be suitable to compare to direct
AdS/dCFT calculations and decide about the parameter $m$ and $k_{i}$.

\section{Asymptotic AdS/dCFT overlaps}

In the AdS/dCFT setting a codimension one defect at $x_{4}=0$ is
introduced in the ${\cal N}=4$ SYM theory, which splits the space
into two halves having $SU(N)$ and $SU(N+k)$ gauge groups. On the
$SU(N+k)$ side the symmetry is broken into $SU(N)$ by the nontrivial
expectation values of three of the six scalar fields:
\begin{equation}
\phi_{i}(x)=-\frac{1}{x_{4}}\text{\ensuremath{\left(\begin{array}{cc}
(S_{i})_{k\times k} & 0_{k\times N}\\
0_{N\times k} & 0_{N\times N}
\end{array}\right)}\ensuremath{\quad;\qquad i=1,2,3}}
\end{equation}
where $S_{i}$ form a $k$-dimensional representation of the $\mathfrak{su}(2)$
algebra. The one-point functions of local single trace gauge-invariant
operators take the form 
\begin{equation}
\langle{\cal O}_{\Delta}(x)\rangle=\frac{c_{\Delta}}{x_{4}^{\Delta}}
\end{equation}
where $\Delta$ is the scaling dimension of the operator. Scaling
operators correspond to Bethe states and at tree level the vacuum
expectation values can be calculated as overlaps with a matrix product
sate
\begin{equation}
c_{\Delta}=\frac{\langle\mathrm{MPS}_{k}\vert\mathbf{u}\rangle}{\sqrt{\langle\mathbf{u}\vert\mathbf{u}\rangle}}\quad;\qquad\langle\mathrm{MPS}_{k}\vert=\sum_{i_{1},\dots,i_{L}}\text{Tr}(S_{i_{1}}\dots S_{i_{L}})\langle\phi_{i_{1}}\vert\otimes\dots\otimes\langle\phi_{i_{L}}\vert
\end{equation}

In \cite{Gombor:2020kgu} we developed a nesting method to calculate
the K-matrices of the excitations in nested Bethe Ansatz systems and
analyzed the cases when the matrix product states were in the two
dimensional representations of $\mathfrak{su}(2)$. Here we extend
those calculations for generic $k$-dimensional representations. Thus
we take generators $S_{1},S_{2},S_{3}$ of the $\mathfrak{su}(2)$
algebra 
\begin{equation}
\left[S_{3},S_{\pm}\right]=\pm S_{\pm}\quad,\qquad\left[S_{+},S_{-}\right]=S_{3}\quad;\qquad S_{\pm}=\frac{1}{\sqrt{2}}\left(S_{1}\pm iS_{2}\right)
\end{equation}
in the $k=2j+1$ dimensional representation where $S_{3}$ acts diagonally:
\begin{equation}
S_{3}=\mathrm{diag}\left(j,j-1,\dots,-j+1,-j\right)
\end{equation}
 and analyze matrix product boundary states with these bond representations.
As a warm up we analyze the $\mathfrak{su}(3)$ spin chain then we
turn to the $\mathfrak{so}(6)$ chain relevant for the D3-D5 brane
setup in the AdS/dCFT correspondence. Before doing so we summarize
our strategy for the calculations by putting special emphasis on the
novel features.

\subsection{Nesting for overlaps with an MPS}

In \cite{Gombor:2020kgu} we developed a nesting method to calculate
the overlaps of a two-site boundary state with periodic Bethe states
of spin chains. Here we generalize this method for the case when the
boundary state is a matrix product state.

Let us summarize the nesting for two-site boundary states
\begin{equation}
\bigl\langle\Psi\bigr|=\psi\otimes\dots\otimes\psi=\psi^{\otimes\frac{L}{2}}
\end{equation}
which satisfy the integrability condition
\begin{equation}
\bigl\langle\Psi\bigr|t(u)=\bigl\langle\Psi\bigr|\Pi t(u)\Pi
\end{equation}
where $t(u)$ and $\Pi t(u)\Pi$ are the transfer matrix and its space
reflected version (see \cite{Piroli:2017sei} for the definition of
integrability). For all of the so far investigated integrable boundary
states the normalized overlap with Bethe states have the following
factorized form
\begin{equation}
\frac{\bigl\langle\Psi\bigr|\mathbf{u}^{(a)}\bigr\rangle}{\sqrt{\bigl\langle\mathbf{u}^{(a)}\bigr|\mathbf{u}^{(a)}\bigr\rangle}}=\prod_{a}\prod_{i=1}^{N_{a}/2}h^{(a)}(u_{i}^{(a)})\sqrt{\frac{\det G_{+}}{\det G_{-}}}\label{eq:exactOverlap}
\end{equation}
where $\Psi$ has a normalized overlap with the pseudovacuum $\langle\Psi\vert0\rangle=1$
and $G_{\pm}$ are the Gaudin-like determinants \cite{Brockmann_2014_1,Brockmann_2014_2,Pozsgay:2018ixm}.
These factorized formulas were verified numerically for various systems
and integrable states \cite{deLeeuw:2015hxa,Buhl-Mortensen:2015gfd,deLeeuw:2016umh,deLeeuw:2018mkd,Pozsgay:2018ixm,Piroli:2018ksf,Piroli:2018don,deLeeuw:2019ebw,Jiang:2019xdz,Gombor:2020kgu}
but exact derivations are available only for rank one cases \cite{Brockmann_2014_1,Brockmann_2014_2,Jiang:2020sdw}.
We emphasize that (\ref{eq:exactOverlap}) is an exact overlap formula
i.e. it holds for any length $L$ and magnon numbers $N_{a}$. The
Gaudin-like determinants depend only on the Bethe state and are well
known, but the boundary state dependent functions $h^{(a)}(u)$ are
not known for general states. The method of \cite{Gombor:2020kgu}
does not try to derive the factorized formula (\ref{eq:exactOverlap})
but, instead, gives an efficient method to determine the unknown functions
$h^{(a)}(u)$ assuming that the exact overlap formula has the above
factorized form.

Our method determines the magnonic overlap functions $h^{(a)}$ recursively
using the nested Bethe ansatz. At each level of the nesting we test
the boundary state with a two particle state and calculate the overlap
in the large volume ($L$) limit. Let us see how it goes. In the first
step of the nesting, the Bethe vectors are constructed over the pseudovacuum
as
\begin{equation}
\left|\mathbf{u}^{(a)}\right\rangle =\sum_{\mathbf{a}^{(1)}}\left|\mathbf{u}^{(1)}\right\rangle _{\mathbf{a}^{(1)}}\mathcal{F}_{\mathbf{a}^{(1)}}(\mathbf{u}^{(a)})
\end{equation}
where the vector $\mathbf{a}^{(1)}=\left\{ a_{1}^{(1)},\dots,a_{N_{1}}^{(1)}\right\} $
labels the first level excitations and $\mathcal{F}_{\mathbf{a}^{(1)}}(\mathbf{u}^{(a)})$
is an eigenvector of the second level transfer matrix. At leading
order in $L$ these first level excitations are orthogonal and are
normalized as
\begin{equation}
\lim_{L\to\infty}{}_{\mathbf{b}^{(1)}}\bigl\langle\mathbf{u}^{(1)}\bigr|\mathbf{u}^{(1)}\bigr\rangle_{\mathbf{a}^{(1)}}=N_{\mathbf{u}^{(1)}}^{2}\delta_{\mathbf{a}^{(1)},\mathbf{b}^{(1)}}+\dots.
\end{equation}
where the dots represent subleading terms in the number of sites,
$L$. The norm of the Bethe vectors in this limit is factorized as
\begin{equation}
\lim_{L\to\infty}\bigl\langle\mathbf{u}^{(a)}\bigr|\mathbf{u}^{(a)}\bigr\rangle=N_{\mathbf{u}^{(1)}}^{2}\sum_{\mathbf{a}^{(1)}}\mathcal{F^{*}}_{\mathbf{a}^{(1)}}(\mathbf{u}^{(a)})\mathcal{F}_{\mathbf{a}^{(1)}}(\mathbf{u}^{(a)})+\dots=:N_{\mathbf{u}^{(1)}}^{2}\bigl\langle\mathbf{u}^{(a)}\bigr|\mathbf{u}^{(a)}\bigr\rangle^{(2)},
\end{equation}
where we introduced the braket notation for the second level Bethe
vectors$\left[\bigr|\mathbf{u}^{(a)}\bigr\rangle^{(2)}\right]_{\mathbf{a}^{(1)}}=\mathcal{F}_{\mathbf{a}^{(1)}}(\mathbf{u}^{(a)}).$
In calculating the normalized overlap
\begin{equation}
\lim_{L\to\infty}\frac{\bigl\langle\Psi\bigr|\mathbf{u}^{(a)}\bigr\rangle}{\sqrt{\bigl\langle\mathbf{u}^{(a)}\bigr|\mathbf{u}^{(a)}\bigr\rangle}}=\frac{\sum_{\mathbf{a}^{(1)}}\bigl\langle\Psi\bigr|\mathbf{u}^{(1)}\bigr\rangle_{\mathbf{a}^{(1)}}\mathcal{F}_{\mathbf{a}^{(1)}}(\mathbf{u}^{(a)})}{N_{\mathbf{u}^{(1)}}\sqrt{\sum_{\mathbf{a}^{(1)}}\mathcal{F^{*}}_{\mathbf{a}^{(1)}}(\mathbf{u}^{(a)})\mathcal{F}_{\mathbf{a}^{(1)}}(\mathbf{u}^{(a)})}}=:\frac{\bigl\langle\Psi(\mathbf{u}^{(1)})\bigr|\mathbf{u}^{(a)}\bigr\rangle^{(2)}}{\sqrt{\bigl\langle\mathbf{u}^{(a)}\bigr|\mathbf{u}^{(a)}\bigr\rangle^{(2)}}},
\end{equation}
we can also introduce the bra notation for the second level boundary
state $\left[^{(2)}\bigl\langle\Psi(\mathbf{u}^{(1)})\bigr|\right]_{\mathbf{a}^{(1)}}=N_{\mathbf{u}^{(1)}}^{-1}\bigl\langle\Psi\bigr|\mathbf{u}^{(1)}\bigr\rangle_{\mathbf{a}^{(1)}}.$
By choosing the normalization of the two-site state as $\bigl\langle\Psi\bigr|0\bigr\rangle=1$
we have ensured a finite $L\to\infty$ limit. Integrability of the
second level boundary state, however, requires an asymptotic factorization
of the form: 
\begin{equation}
\lim_{L\to\infty}{}^{(2)}\bigl\langle\Psi(\mathbf{u}^{(1)})\bigr|=K^{(1)}(u_{1})\otimes\dots\otimes K^{(1)}(u_{N_{1}/2}).\label{eq:conj}
\end{equation}
It is not clear how this follows from the integrability of the state
$\Psi$ and is an assumption in our approach. We validated this assumption
in \cite{Gombor:2020kgu} by calculating already known overlaps with
this method and we saw that the results were consistent. By choosing
the label of the second level pseudo vacuum to be 1, the overlap with
the second level pseudovacuum $\bigr|0\bigr\rangle^{(2)}$ is simply
\begin{equation}
\frac{\bigl\langle\Psi(\mathbf{u}^{(1)})\bigr|0\bigr\rangle^{(2)}}{\sqrt{\bigl\langle0\bigr|0\bigr\rangle^{(2)}}}=\prod_{i=1}^{N_{1}/2}K_{11}^{(1)}(u_{i}^{(1)})
\end{equation}
It is natural to identify the result with the level one overlap function:
\begin{equation}
h^{(1)}(u):=K_{11}^{(1)}(u).
\end{equation}
Since in the known cases the overlap functions $h^{(a)}$ does not
depend on the inhomogeneities we define the second level two site
state as
\begin{align}
^{(2)}\bigl\langle\Psi\bigr| & =\psi^{(2)}\otimes\dots\otimes\psi^{(2)}\quad;\qquad\psi^{(2)}:=\left.\frac{K^{(1)}(u)}{h^{(1)}(u)}\right|_{u=0}
\end{align}
This second level two-site state is normalized as $\langle\Psi\vert0\rangle^{(2)}=1$
and can be used to repeat the nesting procedure at the second level.
Repeating this procedure we can calculate all overlap functions $h^{(a)}(u)$
, recursively. We emphasize that this method uses assumptions like
(\ref{eq:conj}) at all levels of nesting.

Let us finally note that if $\Psi$ were not normalized i.e. the overlap
of the two site state with the pseudo vacuum were $\psi_{11}=A^{2}$
, implying $\bigl\langle\Psi\bigr|0\bigr\rangle=A^{L}\neq1$, then
the asymptotic limit of the second level final state would have this
extra factor: 
\begin{equation}
\lim_{L\to\infty}{}^{(2)}\bigl\langle\Psi(\mathbf{u}^{(1)})\bigr|=A^{L}K^{(1)}(u_{1})\otimes\dots\otimes K^{(1)}(u_{N_{1}/2})
\end{equation}
This factor also shows up in the full overlap formula
\begin{equation}
\frac{\bigl\langle\Psi\bigr|\mathbf{u}^{(a)}\bigr\rangle}{\sqrt{\bigl\langle\mathbf{u}^{(a)}\bigr|\mathbf{u}^{(a)}\bigr\rangle}}=A^{L}\prod_{a}\prod_{i=1}^{N_{a}/2}h^{(a)}(u_{i}^{(a)})\sqrt{\frac{\det G_{+}}{\det G_{-}}}.
\end{equation}

Let us generalize this method to MPS:
\begin{equation}
\bigl\langle\mathrm{MPS}\bigr|=\sum_{i_{1}\dots i_{L}}\mathrm{Tr}\left[\omega_{i_{1}}\dots\omega_{i_{L}}\right]\bigl\langle i_{1}\dots i_{L}\bigr|
\end{equation}
where $\omega_{k}\in\mathrm{End}(V_{B})$, $V_{B}$ is the boundary
vector space. In the calculation above, we did not use the periodic
boundary condition, therefore in the generalization we cut the trace
and define the matrix valued MPS in the boundary space as:
\begin{equation}
_{\alpha\beta}\bigl\langle\mathrm{MPS}\bigr|=\sum_{i_{1}\dots i_{L}}\left(\omega_{i_{1}}\dots\omega_{i_{L}}\right)_{\alpha\beta}\bigl\langle i_{1}\dots i_{L}\bigr|.
\end{equation}
 In calculating the normalized overlap
\begin{equation}
\lim_{L\to\infty}\frac{_{\alpha\beta}\bigl\langle\mathrm{MPS}\bigr|\mathbf{u}^{(a)}\bigr\rangle}{\sqrt{\bigl\langle\mathbf{u}^{(a)}\bigr|\mathbf{u}^{(a)}\bigr\rangle}}=\frac{\sum_{\mathbf{a}^{(1)}}{}_{\alpha\beta}\bigl\langle\mathrm{MPS}\bigr|\mathbf{u}^{(1)}\bigr\rangle_{\mathbf{a}^{(1)}}\mathcal{F}_{\mathbf{a}^{(1)}}(\mathbf{u}^{(a)})}{N_{\mathbf{u}^{(1)}}\sqrt{\sum_{\mathbf{a}^{(1)}}\mathcal{F^{*}}_{\mathbf{a}^{(1)}}(\mathbf{u}^{(a)})\mathcal{F}_{\mathbf{a}^{(1)}}(\mathbf{u}^{(a)})}}=\frac{_{\alpha\beta}\bigl\langle\Psi(\mathbf{u}^{(1)})\bigr|\mathbf{u}^{(a)}\bigr\rangle^{(2)}}{\sqrt{\bigl\langle\mathbf{u}^{(a)}\bigr|\mathbf{u}^{(a)}\bigr\rangle^{(2)}}},
\end{equation}
 we also introduce the bra notation for the second level final state
\begin{equation}
\left[_{\alpha\beta}^{(2)}\bigl\langle\Psi(\mathbf{u}^{(1)})\bigr|\right]_{\mathbf{a}^{(1)}}=\frac{_{\alpha\beta}\bigl\langle\mathrm{MPS}\bigr|\mathbf{u}^{(1)}\bigr\rangle_{\mathbf{a}^{(1)}}}{N_{\mathbf{u}^{(1)}}}.
\end{equation}
We have to calculate this overlap for all $\alpha,\beta$. It is obvious,
however, that all components can not be normalized as $_{\alpha\beta}\bigl\langle\mathrm{MPS}\bigr|0\bigr\rangle=1$
simultaneously. This implies $A^{L}$-like factors even at the asymptotic
limit, which has to be taken carefully.

Usually, the asymptotic second level final state is not irreducible
in the boundary space, rather it can be decomposed into irreducible
components as 
\begin{equation}
\lim_{L\to\infty}{}_{\alpha\beta}^{(2)}\bigl\langle\Psi(\mathbf{u}^{(1)})\bigr|=\left(\bigl\langle\Psi_{1}(\mathbf{u}^{(1)})\bigr|\oplus_{V_{B}}\bigl\langle\Psi_{2}(\mathbf{u}^{(1)})\bigr|\dots\oplus_{V_{B}}\bigl\langle\Psi_{n}(\mathbf{u}^{(1)})\bigr|\right)_{\alpha\beta}
\end{equation}
where $\dim_{V_{B}}\left(\bigl\langle\Psi_{i}(\mathbf{u}^{(1)})\bigr|\right)=d_{i}$.
We assume that the matrix part is factorized into a product of two-particle
(matrix valued) K-matrices in the asymptotic limit, i.e., $\bigl\langle\Psi_{m}(\mathbf{u}^{(1)})\bigr|$
can be written as 
\begin{equation}
_{\alpha\beta}\bigl\langle\Psi_{m}(\mathbf{u}^{(1)})\bigr|=A_{m}^{L}K_{\alpha\gamma_{1}}^{(1,m)}(u_{1}^{(1)})\otimes K_{\gamma_{1}\gamma_{2}}^{(1,m)}(u_{2}^{(1)})\otimes\dots\otimes K_{\gamma_{N_{1}/2}\beta}^{(1,m)}(u_{N_{1}/2}^{(1)})
\end{equation}
where the excitation K-matrices $K_{\alpha\beta}^{(1,m)}(u)$ do not
depend on $L$. Applying this formula for the pseudovacuum ($\mathbf{u}^{(1)}=\left\{ \right\} $)
gives:
\begin{equation}
_{\alpha\beta}\bigl\langle\Psi_{m}(\left\{ \right\} )\bigr|=A_{m}^{L}\delta_{\alpha\beta}
\end{equation}
therefore
\begin{equation}
_{\alpha\beta}\bigl\langle\mathrm{MPS}\bigr|0\bigr\rangle=\left(A_{1}^{L}\mathbf{1}_{d_{1}}\oplus_{V_{B}}A_{2}^{L}\mathbf{1}_{d_{2}}\oplus_{V_{B}}\dots\oplus_{V_{B}}A_{n}^{L}\mathbf{1}_{d_{n}}\right)_{\alpha\beta}.
\end{equation}
This equation implies that we can obtain $A_{k}$s from the vacuum
overlap and it suggests the irreducible decomposition.

In the following we focus on the case when the irreducible second
level states are one dimensional:
\begin{equation}
\bigl\langle\Psi_{m}(\mathbf{u}^{(1)})\bigr|=A_{m}^{L}K^{(1,m)}(u_{1}^{(1)})\otimes K^{(1,m)}(u_{2}^{(1)})\otimes\dots\otimes K^{(1,m)}(u_{N_{1}/2}^{(1)}),
\end{equation}
such that we can apply the previous method for each one dimensional
part. This amounts to define first level one particle functions and
second level two-site states as 
\begin{align}
h_{m}^{(1)}(u) & :=K_{11}^{(1,m)}(u)\quad;\qquad\psi_{m}^{(2)}:=\left.\frac{K^{(1,m)}(u)}{h_{m}^{(1)}(u)}\right|_{u=0}
\end{align}
We can now iteratively calculate the one particle overlaps for each
irreducible piece and the proposed final overlap takes the form 
\begin{equation}
\frac{\bigl\langle\mathrm{MPS}\bigr|\mathbf{u}^{(a)}\bigr\rangle}{\sqrt{\bigl\langle\mathbf{u}^{(a)}\bigr|\mathbf{u}^{(a)}\bigr\rangle}}=\sum_{m}A_{m}^{L}\prod_{a}\prod_{i=1}^{N_{a}/2}h_{m}^{(a)}(u_{i}^{(a)})\sqrt{\frac{\det G_{+}}{\det G_{-}}}.
\end{equation}

For the justification of the used assumptions we carry out this procedure
for certain MPS in the $\mathfrak{su}(3)$ and $\mathfrak{so}(6)$
spin chains for which the overlaps are already known.

\subsection{$\mathfrak{su}(3)$ spin chain}

Let us consider the following matrix product state
\begin{equation}
\left\langle \mathrm{MPS}_{k}\right|=\mathrm{Tr}\left[\left(S_{1}\left\langle 1\right|+S_{2}\left\langle 2\right|+S_{3}\left\langle 3\right|\right)^{\otimes L}\right]
\end{equation}
and calculate its overlap in the $\mathfrak{su}(3)$ spin chain with
the eigenvectors of the periodic transfer matrix. The excitation K-matrix
can be extracted from the large $L$ limit of the overlap between
the MPS and a generic two particle state built over the pseudo vacuum
which we choose to be $\bigr|0\bigr\rangle=\bigr|3\bigr\rangle^{\otimes L}$.

At first let us calculate overlap of the pseudovacuum:
\begin{equation}
_{\alpha\beta}\bigl\langle\mathrm{MPS}_{k}\bigr|0\bigr\rangle=\mathrm{diag}\left(j^{L},(j-1)^{L},\dots,(-j+1)^{L},(-j)^{L}\right)_{\alpha\beta}
\end{equation}
This vacuum overlap suggest that the second level final state is a
direct sum of one dimensional states. In coordinate space Bethe ansatz
the two particle plane wave takes the form 
\begin{equation}
\left|u,-u\right\rangle _{ab}=\sum_{n_{1}=1}^{L}\sum_{n_{2}=n_{1}+1}^{L}\left(e^{ip(n_{1}-n_{2})}\left|n_{1}n_{2}\right\rangle _{ab}+e^{-ip(n_{1}-n_{2})}S_{ab}^{cd}(2u)\left|n_{1}n_{2}\right\rangle _{cd}\right).
\end{equation}
where $\vert nm\rangle_{ab}$ denotes a state with excitations $a=1,2$
at position $n$ and $b=1,2$ at $m$ and their $\mathfrak{su}(2)$
invariant S-matrix is normalized as $S_{11}^{11}(u)=1$ and so $S_{12}^{21}(u)=\frac{u+i}{u-i}$.
The overlap in the large $L$ limit can be written as
\begin{equation}
\lim_{L\to\infty}\,\frac{_{\alpha\beta}\Bigl\langle\mathrm{MPS}_{k}\Bigr|u,-u\Bigr\rangle_{ab}}{N_{u,-u}}=(\Sigma_{ab}(u)+S_{ab}^{cd}(2u)\Sigma_{cd}(-u))_{\alpha\beta}
\end{equation}
 where
\begin{equation}
\Sigma_{ab}(u)=\lim_{L\to\infty}\frac{1}{L}\sum_{n_{1}=1}^{L}\sum_{n_{2}=n_{1}+1}^{L}e^{ip(n_{1}-n_{2})}S_{3}^{n_{1}-1}S_{a}S_{3}^{n_{2}-n_{1}-1}S_{b}S_{3}^{L-n_{2}}.
\end{equation}
By changing the basis from $a,b=1,2$ to $A,B=+,-$ and using the
commutation relations we can calculate 
\begin{equation}
\Sigma_{AB}(u)=\lim_{L\to\infty}\frac{1}{L}\sum_{n_{1}<n_{2}}e^{ip(n_{1}-n_{2})}S_{A}S_{B}(S_{3}+A+B)^{n_{1}-1}(S_{3}+B)^{n_{2}-n_{1}-1}S_{3}^{L-n_{2}}
\end{equation}
We can then see that $\Sigma_{\pm\mp}(u)=\text{diag}(\Sigma_{\pm\mp}^{j,j}(u),\Sigma_{\pm\mp}^{j,j-1}(u),\dots,\Sigma_{\pm\mp}^{j,-j+1}(u),\Sigma_{\pm\mp}^{j,-j}(u))$
are diagonal matrices with entries 
\begin{align}
\Sigma_{\pm\mp}^{j,m}(u) & =\frac{1}{2}\left(j(j+1)-m(m\mp1)\right)\frac{1}{m(e^{ip}-1)\pm1},
\end{align}
while the non-diagonal $\Sigma_{++}(u)$ and $\Sigma_{--}(u)$ components
do not have any $O(L)$ piece, thus they vanish in the $L\to\infty$
limit. In calculating the matrix element $\Sigma_{\pm\mp}^{j,m}$
we have to normalize the overlap with the pseudo vacuum to $1$. This
amounts to renormalizing the overlaps with the factor $m^{L}$ and
introduces different normalization for the various diagonal components
of the overlap
\begin{align}
\lim_{L\to\infty}\frac{_{\alpha\beta}\Bigl\langle\mathrm{MPS}_{k}\Bigr|u,-u\Bigr\rangle_{ab}}{N_{u,-u}}=\,\,\,\,\,\,\,\,\,\,\,\,\,\,\,\,\,\,\,\,\,\,\,\,\,\,\,\,\,\,\,\,\,\,\,\,\,\,\,\,\,\,\,\,\,\,\,\,\,\,\,\,\,\,\,\,\,\,\,\,\,\,\,\,\,\,\,\,\,\,\,\,\,\,\,\,\,\,\,\,\,\,\,\,\,\,\,\,\,\,\,\,\,\,\,\,\,\,\,\,\,\,\,\,\,\,\,\,\,\,\,\,\,\,\,\,\,\,\,\,\,\,\,\,\,\,\,\,\,\,\,\,\,\,\,\,\,\,\,\,\,\,\,\,\,\,\,\,\,\,\,\,\,\,\,\,\,\,\,\,\,\,\,\,\,\,\,\,\,\\
\mathrm{diag}\left(j^{L}K_{ab}^{(j,j)}(u),(j-1)^{L}K_{ab}^{(j,j-1)}(u),\dots,(-j+1)^{L}K_{ab}^{(j,-j+1)}(u),(-j)^{L}K_{ab}^{(j,-j)}(u)\right)_{\alpha\beta}\nonumber 
\end{align}
where the diagonal scalar K-matrices of the excitations take the form
\begin{equation}
K_{ab}^{(j,m)}(u)=\frac{u\left(u+i/2\right)\left(u^{2}+\left(j+1/2\right)^{2}\right)}{\left(u^{2}+\left(m+1/2\right)^{2}\right)\left(u^{2}+\left(m-1/2\right)^{2}\right)}\left(\begin{array}{cc}
1 & -\frac{m}{u+i/2}\\
\frac{m}{u+i/2} & 1
\end{array}\right)\label{eq:exK}
\end{equation}
which means that the first level one-particle overlaps are
\begin{equation}
h_{m}^{(1)}(u)=\frac{u\left(u+i/2\right)\left(u^{2}+\left(j+1/2\right)^{2}\right)}{\left(u^{2}+\left(m+1/2\right)^{2}\right)\left(u^{2}+\left(m-1/2\right)^{2}\right)}
\end{equation}
and the irreducible second level $K$-matrices are
\begin{equation}
\left(\begin{array}{cc}
1 & -\frac{m}{u+i/2}\\
\frac{m}{u+i/2} & 1
\end{array}\right).
\end{equation}
Now we can use the known $\mathfrak{su}(2)$ boundary state overlaps
of these $K$-matrices \cite{Gombor:2020kgu} 
\begin{equation}
h_{m}^{(2)}(u)=\frac{\left(u^{2}+m^{2}\right)}{u\left(u+i/2\right)}
\end{equation}
to make a proposition for the full overlap. In doing so we have to
restore the relative normalization of the various diagonal contributions
\begin{multline}
\frac{\Bigl\langle\mathrm{MPS}_{k}\Bigr|\mathbf{u},\mathbf{v}\Bigr\rangle}{\sqrt{\Bigl\langle\mathbf{u},\mathbf{v}\Bigr|\mathbf{u},\mathbf{v}\Bigr\rangle}}=\sum_{m=-j}^{j}m^{L}\prod_{i=1}^{N}h_{m}^{(1)}(u_{i})\prod_{i=1}^{M}h_{m}^{(2)}(v_{i})\sqrt{\frac{\det G_{+}}{\det G_{-}}}=\\
\sum_{m=-j}^{j}m^{L}\prod_{i=1}^{N}\frac{u_{i}\left(u_{i}+i/2\right)\left(u_{i}^{2}+\left(j+1/2\right)^{2}\right)}{\left(u_{i}^{2}+\left(m+1/2\right)^{2}\right)\left(u_{i}^{2}+\left(m-1/2\right)^{2}\right)}\prod_{i=1}^{M}\frac{\left(v_{i}^{2}+m^{2}\right)}{v_{i}\left(v_{i}+i/2\right)}\sqrt{\frac{\det G_{+}}{\det G_{-}}}=\\
\sum_{m=-j}^{j}m^{L}\frac{Q_{1}\left(ij+i/2\right)Q_{2}\left(im\right)}{Q_{1}\left(im+i/2\right)Q_{1}\left(im-i/2\right)}\sqrt{\frac{Q_{1}(0)Q_{1}\left(i/2\right)}{Q_{2}(0)Q_{2}\left(i/2\right)}\frac{\det G_{+}}{\det G_{-}}}\label{eq:su3}
\end{multline}
where we used the $Q$-functions
\begin{equation}
Q_{1}(u)=\prod_{i=1}^{N}(u-u_{i}),\qquad Q_{2}(u)=\prod_{i=1}^{N}(u-v_{i}).
\end{equation}
We can see that (\ref{eq:su3}) agrees with the numerically proved
formula in \cite{deLeeuw:2016umh,deLeeuw:2018mkd}, which verifies
that the excitation K-matrix (\ref{eq:exK}) we just calculated is
physically meaningful.

\subsection{$\mathfrak{so}(6)$ spin chain}

Let us consider the following MPS
\begin{equation}
^{\alpha,\beta}\left\langle \mathrm{MPS}_{k}\right|=\left[\left(\sqrt{2}\phi_{1}S_{1}+\sqrt{2}\phi_{3}S_{2}+\sqrt{2}\phi_{5}S_{3}\right)^{\otimes L}\right]^{\alpha,\beta}=\left[\left(\sqrt{2}\phi_{1}S_{1}+\sqrt{2}\phi_{3}S_{2}+(Z+\bar{Z})S_{3}\right)^{\otimes L}\right]^{\alpha,\beta}
\end{equation}
Following the conventions and calculations in \cite{Gombor:2020kgu}
we choose the pseudo vacuum as $Z^{\otimes L}$ and label $\mathfrak{so}(4)$
excitations with $a,b=1,\dots,4$. The overlap calculation is similar
to the $\mathfrak{su}(3)$ case the only differences are in the building
block
\begin{equation}
\Sigma_{ab}^{j,m}(u)=\frac{1}{2}\left(\begin{array}{cccc}
\Sigma_{+-}^{j,m}+\Sigma_{-+}^{j,m} & 0 & i\Sigma_{+-}^{j,m}-i\Sigma_{-+}^{j,m} & 0\\
0 & 0 & 0 & 0\\
-i\Sigma_{+-}^{j,m}+i\Sigma_{-+}^{j,m} & 0 & \Sigma_{+-}^{j,m}+\Sigma_{-+}^{j,m} & 0\\
0 & 0 & 0 & 0
\end{array}\right)_{ab}
\end{equation}
from which the diagonal $K$-matrix $(K_{ab}^{(j,j)}(u),K_{ab}^{(j,j-1)}(u),\dots,K_{ab}^{(j,-j+1)}(u),K_{ab}^{(j,-j)}(u))$
can be built up as 
\begin{multline}
K_{ab}^{(j,m)}(u)=\Sigma_{ab}^{j,m}(u)+S_{ab}^{cd}(2u)\Sigma_{cd}^{j,m}(-u)+\delta_{ab}e(2u)\cong\\
\frac{u\left(u+i/2\right)\left(u^{2}+\left(j+1/2\right)^{2}\right)}{\left(u^{2}+\left(m+1/2\right)^{2}\right)\left(u^{2}+\left(m-1/2\right)^{2}\right)}\left(\begin{array}{cc}
1 & \frac{m}{u+i/2}\\
-\frac{m}{u+i/2} & 1
\end{array}\right)\otimes\left(\begin{array}{cc}
1 & \frac{m}{u+i/2}\\
-\frac{m}{u+i/2} & 1
\end{array}\right).\label{eq:so6K}
\end{multline}
Here we also showed how the $\mathfrak{so}(4)$ $K$-matrix can be
factorized into two $\mathfrak{su}(2)$ factors. Based on the $\mathfrak{su}(2)$
overlaps our proposition for the full overlap is
\begin{multline}
\frac{\Bigl\langle\mathrm{MPS}_{k}\Bigr|\mathbf{u},\mathbf{v},\mathbf{w}\Bigr\rangle}{\sqrt{\Bigl\langle\mathbf{u},\mathbf{v},\mathbf{w}\Bigr|\mathbf{u},\mathbf{v},\mathbf{w}\Bigr\rangle}}=\\
\sum_{m=-j}^{j}m^{L}\prod_{i=1}^{N}\frac{u_{i}(u_{i}+\frac{i}{2})(u_{i}^{2}+\frac{(2j+1)^{2}}{4})}{(u_{i}^{2}+\frac{(2m-1)^{2}}{4})(u_{i}^{2}+\frac{(2m+1)^{2}}{4})}\prod_{i=1}^{M_{+}}\frac{v_{i}^{2}+m^{2}}{v_{i}(v_{i}+\frac{i}{2})}\prod_{i=1}^{M_{-}}\frac{w_{i}^{2}+m^{2}}{w_{i}(w_{i}+\frac{i}{2})}\sqrt{\frac{\det G_{+}}{\det G_{-}}}=\\
\sum_{m=-j}^{j}m^{L}\prod_{i=1}^{N}\frac{Q_{1}\left(ij+i/2\right)Q_{+}\left(im\right)Q_{-}\left(im\right)}{Q_{1}\left(im+i/2\right)Q_{1}\left(im-i/2\right)}\sqrt{\frac{Q_{1}(0)Q_{1}\left(i/2\right)}{Q_{+}(0)Q_{+}\left(i/2\right)Q_{-}(0)Q_{-}\left(i/2\right)}\frac{\det G_{+}}{\det G_{-}}}\label{eq:so6}
\end{multline}
which again agrees with the numerically proved formula \cite{deLeeuw:2018mkd}.
In (\ref{eq:so6}) we used the $\mathfrak{so}(6)$ $Q$-functions
\begin{equation}
Q_{1}(u)=\prod_{i=1}^{N}(u-u_{i}),\qquad Q_{+}(u)=\prod_{i=1}^{N}(u-v_{i}),\qquad Q_{-}(u)=\prod_{i=1}^{N}(u-w_{i}).
\end{equation}

\subsection{Proposal for the asymptotic overlap in AdS/dCFT for generic k}

Comparing the $\mathfrak{so}(6)$ result (\ref{eq:so6K}) with the
weak coupling limit of section \ref{subsec:Weak-coupling-expansion}
it is natural to assume a diagonal K-matrix structure in the boundary
space consisting of the diagonal elements 
\begin{equation}
\mathbb{K}^{AdS/dCFT}=\mathrm{diag}\left(\mathbb{K}^{(j,j)}(u),\mathbb{K}^{(j,j-1)}(u),\dots,\mathbb{K}^{(j,1-j)}(u),\mathbb{K}^{(j,-j)}(u)\right)
\end{equation}
where $\mathbb{K}^{(j,m)}(u)$ is defined in (\ref{eq:Kjm}) and the
parameters are fixed as $k_{1}=k_{4}=1$ and $k_{2}=0$. Thus the
one-point function in the asymptotic limit can be calculated as the
overlap of the boundary state $B_{k}$ based on this K-matrix and
the periodic multi-particle state
\begin{align}
c_{\Delta} & =\frac{\Bigl\langle\mathrm{B}_{k}\Bigr|\mathbf{u},\mathbf{y}^{(\nu)},\mathbf{w}^{(\nu)}\Bigr\rangle}{\sqrt{\Bigl\langle\mathbf{u},\mathbf{y}^{(\nu)},\mathbf{w}^{(\nu)}\Bigr|\mathbf{u},\mathbf{y}^{(\nu)},\mathbf{w}^{(\nu)}\Bigr\rangle}}=\sum_{m=-j}^{j}\frac{x_{m}^{L}\Bigl\langle\mathrm{B}^{(j,m)}\Bigr|\mathbf{u},\mathbf{y}^{(\nu)},\mathbf{w}^{(\nu)}\Bigr\rangle}{\sqrt{\Bigl\langle\mathbf{u},\mathbf{y}^{(\nu)},\mathbf{w}^{(\nu)}\Bigr|\mathbf{u},\mathbf{y}^{(\nu)},\mathbf{w}^{(\nu)}\Bigr\rangle}}\label{eq:overlap}
\end{align}
where the Bethe states are labeled by the Bethe roots $\mathbf{y}^{(\nu)},\mathbf{w}^{(\nu)}$
with $\nu=1,2$ of the two $\mathfrak{su}(2\vert2)$ wings. The matrix
product type boundary state involves a trace coming from the boundary
inner degree of freedom. Since the K-matrix is diagonal the overlap
is simply the sum of individual overlaps with $\langle B^{(j,m)}\vert$
, which is the boundary state built from the scalar (in boundary space)
K-matrix $\mathbb{K}^{(j,m)}.$ To decide how to normalize the individual
components we used the results of \cite{Komatsu:2020sup}, which was
obtained from exact localization calculations and also from bootstrap
considerations and was checked against perturbative calculations \cite{Buhl-Mortensen:2016pxs}.
Fortunately we have already calculated the overlap (\ref{eq:overlap})
in \cite{Gombor:2020kgu} thus the full result we propose for the
one point function is 
\begin{equation}
c_{\Delta}=\sum_{m=-j}^{j}x_{m}^{L}\prod_{i=1}^{N/2}K_{j,m}(u_{i})\prod_{\nu=1}^{2}\prod_{i=1}^{M_{1}^{(\nu)}/2}\frac{(y_{i}^{(\nu)})^{2}+x_{m}^{2}}{ix_{m}y_{i}^{(\nu)}}\prod_{i=1}^{\left\lfloor M_{1}^{(\nu)}/4\right\rfloor }\frac{1}{w_{i}^{(\nu)}\left(w_{i}^{(\nu)}+\frac{i}{2g}\right)}\sqrt{\frac{\det G_{+}}{\det G_{-}}}
\end{equation}
where the Gaudin like determinants can be calculated from the BA equations
\cite{Gombor:2020kgu}.

\section{Conclusions}

We formulated and analyzed the boundary state bootstrap for factorizing
K-matrices with residual $\mathfrak{osp}(2\vert2)_{c}$ symmetry.
This K-matrix had three ``bosonic'' and one ``fermionic'' orientations
parameterized by $k_{i}$ and $x_{s}$, respectively. By bootstrapping
on the bound state pole we showed that the matrix structure of the
excited boundary state had the same form as the ground state one,
merely the parameter $x_{s}$ was shifted as $x_{s\pm1}$. We then
analyzed the pole structure of the excited boundary K-matrix and identified
a new boundary bound state. For semi integer $s$ we closed the boundary
bootstrap by providing the full set of boundary bound states together
with their K-matrices and explanations of all of their poles. In order
to apply these results for higher dimensional MPS in the D3-D5 AdS/dCFT
setting we calculated the tree level excitation K-matrix. We did it
first in an $\mathfrak{su}(3)$ spin chain, then we extended the result
for the $\mathfrak{so}(6)$ spin chain. In both cases the excitation
K-matrices were diagonal in the boundary space without any inner degree
of freedom. By generalizing our nesting method developed in \cite{Gombor:2020kgu}
for matrix product states we calculated the most general overlaps
at tree level. These agreed with previous results in the literature
\cite{deLeeuw:2016umh,deLeeuw:2018mkd}. The comparison of the tree
level excitation K-matrix with the weak coupling limit of our all
loop K-matrix fixed its parameters. Based on our previous results
we could propose the all loop asymptotic overlaps for the D3-D5 system
including all excitations. These extend the results of \cite{Komatsu:2020sup}
from the diagonal $\mathfrak{su}(2)$ sector for the full theory.

Our results are asymptotic in the sense that it does not contain any
wrapping corrections. To incorporate those effects of the virtual
particles one has to go through the program suggested in \cite{Komatsu:2020sup,Jiang:2019xdz,Jiang:2019zig}
The recent results \cite{Kostov:2018dmi,Kostov:2019fvw,Kostov:2019sgu,Caetano:2020dyp}
on excited state g-functions could help to incorporate the wrapping
corrections also in our case, while the result \cite{Caetano:2020dyp}
might sheds some light how it could be formulated more efficiently.

There are two ways to embed the residual $\mathfrak{osp}(2\vert2)_{c}$
symmetry into $\mathfrak{su}(2\vert2)_{c}$ depending on which $\mathfrak{su}(2)$
factor is unbroken. Here we analyzed the case with the full Lorentzian
symmetry. Completely analogous considerations can be performed for
the case when the R-symmetry is unbroken. Those results can be relevant
for correlation functions with Wilson loops.

\subsection*{Acknowledgments}

We thank the NKFIH research Grants K116505 and K134946 for support.

\bibliographystyle{elsarticle-num}
\bibliography{refs}

\end{document}